\documentclass[fleqn,usenatbib]{mnras}

\usepackage{newtxtext,newtxmath}

\usepackage[T1]{fontenc}

\DeclareRobustCommand{\VAN}[3]{#2}
\let\VANthebibliography\thebibliography
\def\thebibliography{\DeclareRobustCommand{\VAN}[3]{##3}\VANthebibliography}

\usepackage{graphicx}	
\usepackage{amsmath}	
\usepackage{siunitx}
\usepackage{pifont}

\newcommand{\oth}{O\,{\sc iii}}

\defcitealias{Andrews2013}{AM13}

\title[Correcting $T_e$-method metallicity indicators]{A novel approach to correcting $T_e$-based mass-metallicity relations}

\author[A. J. Cameron et al.] {Alex J. Cameron,\thanks{E-mail:
  \href{mailto:alex.cameron@physics.ox.ac.uk}{alex.cameron@physics.ox.ac.uk}} Harley Katz, and Martin P. Rey
  \\
  Sub-department of Astrophysics, University of Oxford, Keble Road, Oxford OX1 3RH, United Kingdom
  }

\date{Accepted XXX. Received YYY; in original form ZZZ}

\pubyear{2022}

\begin{document}
\label{firstpage}
\pagerange{\pageref{firstpage}--\pageref{lastpage}}
\maketitle

\begin{abstract}
Deriving oxygen abundances from the electron temperature (hereafter the $T_e$-method) is the gold-standard for extragalactic metallicity studies. However, unresolved temperature fluctuations within individual H~{\sc ii} regions and across different H~{\sc ii} regions throughout a galaxy can bias metallicity estimates low, with a magnitude that depends on the underlying and typically unknown temperature distribution. Using a toy model, we confirm that computing $T_e$-based metallicities using the temperature derived from the [O {\sc iii}] $\lambda$4363/$\lambda$5007 or [O~{\sc ii}]~$\lambda\lambda$7320,7330 / [O~{\sc ii}]~$\lambda\lambda$3727 ratio (`ratio temperature'; $T_{\rm ratio}$) results in an underprediction of metallicity when temperature fluctuations are present. In contrast, using the unobservable `line temperatures' ($T_{\rm line}$) that provide the mean electron and ion density-weighted emissivity yield an accurate metallicity estimate. To correct this bias in low-mass galaxies, we demonstrate an example calibration of a relation between $T_{\rm ratio}$ and $T_{\rm line}$ based on a high-resolution (4.5~pc) {\small RAMSES-RTZ} simulation of a dwarf galaxy that self-consistently models the formation of multiple H~{\sc ii} regions and ion temperature distribution in a galactic context. Applying this correction to the low-mass end of the mass-metallicity relation shifts its normalization up by 0.18~dex on average and flattens its slope from 0.87 to 0.58, highlighting the need for future studies to account for, and correct, this bias.
\end{abstract}

\begin{keywords}
ISM: abundances -- galaxies: abundances -- galaxies: HII regions -- galaxies: ISM -- galaxies: evolution
\end{keywords}



\section{Introduction}
\label{sec:intro}

\begin{figure*}
\centerline{
\includegraphics[scale=1.0,trim={0 1cm 0cm 1cm},clip]{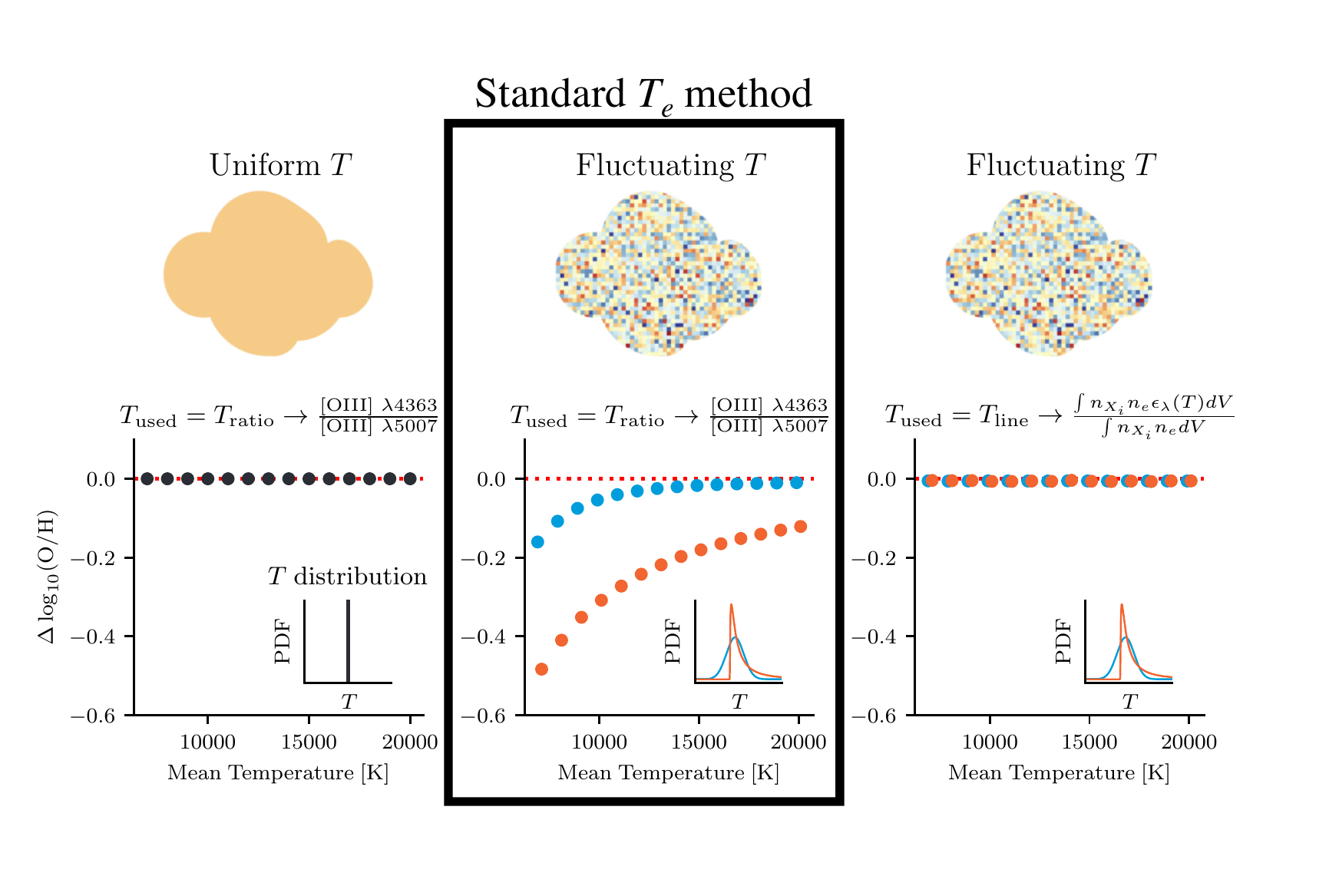}}
\caption{
Toy-model demonstrating how temperature fluctuations bias the $T_e$-method when applied to a uniform metallicity system. From left to right, lower panels show the discrepancy between the true metallicity and that derived from Equation~\ref{eqn:DM} for: a homogeneous cloud with uniform temperature and using $T_{\rm ratio}$ (the $T_e$ inferred from [\oth] $\lambda$4363/$\lambda$5007), a cloud with temperature fluctuations and using $T_{\rm ratio}$, and a cloud with temperature fluctuations and using $T_{\rm line}$ (Equation~\ref{eqn:t_line}). Temperature fluctuations are either normally (blue) or lognormally (orange) distributed (insets). The standard $T_e$-method (centre panel) always underpredicts metallicity when temperature fluctuations are present, with the deficit depending on the exact shape of the temperature distribution. In contrast, when the cloud is homogeneous or when $T_{\rm line}$ is used, the metallicity prediction is accurate.
}
\label{fig:temp_sims}
\end{figure*}

The gas-phase oxygen abundance (`metallicity' hereafter) of a galaxy is a fundamental property that encodes a wealth of information on its star formation, enrichment, and assembly history. Galaxy metallicity exhibits a tight correlation with stellar mass, the mass-metallicity relation (MZR; e.g. \citealt{Lequeux1979}, \citealt{Tremonti2004}), with scatter in this relation correlating with star formation rate \citep[e.g.][]{Ellison2008, Mannucci2010} and gas mass \citep[e.g.][]{Bothwell2013}. Measurements of the slope and normalisation of this relation are interpreted as the collective effect of metal-enriched outflows and metal-poor inflows \citep[e.g.][]{Larson1974,Dalcanton2004,Tremonti2004}, and used to constrain astrophysical models that predict the chemical enrichment and assembly histories of galaxies \citep[e.g.][]{Ma2016, DeRossi2017, Torrey2019}. 

However, the exact shape of the MZR remains debated, as its slope and normalisation is sensitive to the technique used to measure metallicity \citep[e.g.][]{Kewley2008, Curti2020,Yates2020}. Using recombination lines (RL) is often considered the most robust metallicity estimate, due to its weak dependence on gas temperature and density. But it relies on measuring multiple faint metal lines, which is possible in the local neighbourhood of the Milky Way \citep[e.g.][]{Peimbert1993,Esteban2002} but prohibitive for extragalactic samples. In contrast, strong-line methods rely only on ratios of the brightest emission lines and can therefore be widely applied. However, they need to be calibrated to more direct measurements \citep[e.g.][]{Curti2017} or theoretical predictions \citep[e.g.][]{McGaugh1991, Kewley2019}, each associated with dificult-to-quantify systematic uncertainties \citep{Kewley2008}.

As a result, the $T_e$ method \citep{Peimbert1967} has been the {\it de facto} gold standard for measurements of the MZR (e.g. \citealt{Andrews2013}, hereafter \citetalias{Andrews2013}; \citealt{Curti2020}). It relies on detecting auroral emission lines (e.g. [O {\sc iii}]~$\lambda$4363) to estimate $T_e$, enabling us to derive metallicities solely from line emissivities based on collision strengths from quantum mechanical calculations \citep[e.g.][]{Aggarwal1999,Palay2012,Storey2014}. Detecting such faint auroral lines is observationally challenging, but possible at low-redshift for large samples of individual low-metallicity galaxies \citep[e.g.][]{Izotov2019} or stacks of higher-metallicity galaxies \citep{Curti2017}. Further, the advent of \emph{JWST} now allows us to make such detections at high-redshift (e.g. at $z=8.5$; \citealt{Katz2022-jwst, Curti2022}). The central role played by the $T_e$-method to constrain chemical evolution across cosmic time makes it paramount to pinpoint its potential biases and losses of accuracy. 

In fact, $T_e$-method metallicities are typically lower than RL techniques by as much as $\sim0.3$~dex in observations where both can be computed \citep[e.g.][]{Liu2001,Garcia2007}.
A promising explanation for these offsets is the presence of inhomogeneous temperatures for the emission-line emitting gas, which then biases the $T_e$-method metallicity measurements low due to the exponential scaling of line emissivities with temperature (\citealt{Peimbert1967,Stasinska2005,Bresolin2008, Pilyugin2012}; Section~\ref{sec:simple}). Such temperature inhomogeneities can originate within individual H~{\sc ii} regions due to their internal structure (see  \citealt{Peimbert2019} for a review), but also across a galaxy, as the multiple H~{\sc ii} regions integrated into a galactic spectrum are subject to a diversity of ISM densities, temperatures and feedback conditions. Although $T_e$-metallicities can be debiased by measuring multiple temperatures for the same system to estimate the underlying distribution (\citealt{Peimbert1969, Peimbert2004}), such observations are typically restricted to nearby, individual H~{\sc ii} regions and are rarely feasible for distant galaxies. 

Pinpointing the importance of galactic-scale temperature fluctuations is thus essential to calibrate the true magnitude of this long-standing bias and recover accurate extragalactic $T_e$-metallicities. In this paper, we introduce a new physically-motivated model to achieve this aim. Rather than estimating temperature fluctuations from measurements of local H~{\sc ii} regions \citep[e.g.][]{Kobulnicky1999} or photoionization models \citep{Stasinska1990,Garnett1992}, we use a high-resolution ($\approx 4.5$ pc) simulation of an isolated dwarf galaxy with multifrequency radiation-hydrodynamics coupled to non-equilibrium metal and molecular chemistry. This allows us to self-consistently predict the distribution of H~{\sc ii} region temperatures, across an entire galaxy undergoing star formation and feedback episodes. We estimate the resulting bias in $T_e$ metallicities (Section~\ref{sec:results}) and how it affects the derived MZR (Section~\ref{sec:conclusions}).

\begin{figure}
\centerline{
\includegraphics[scale=1,trim={0 0.cm 0cm 0.cm},clip]{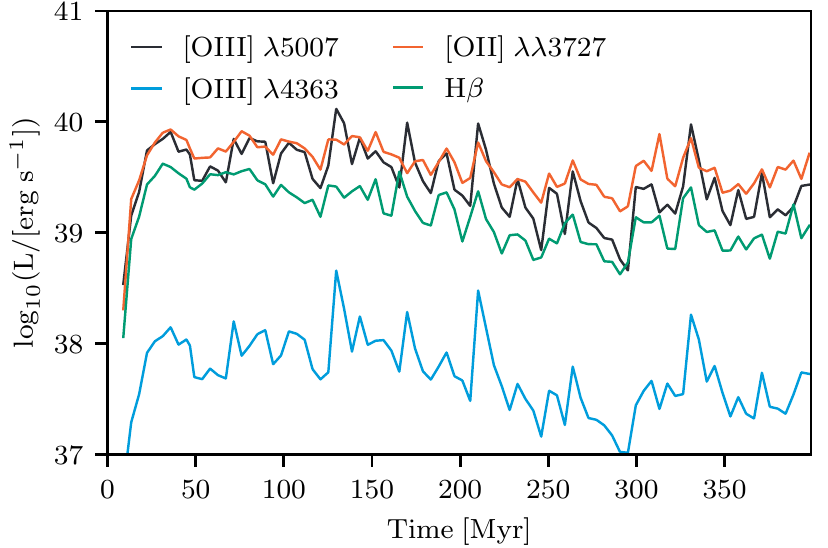}
}
\centerline{
\includegraphics[scale=1,trim={0 0.0cm 0cm 0.0cm},clip]{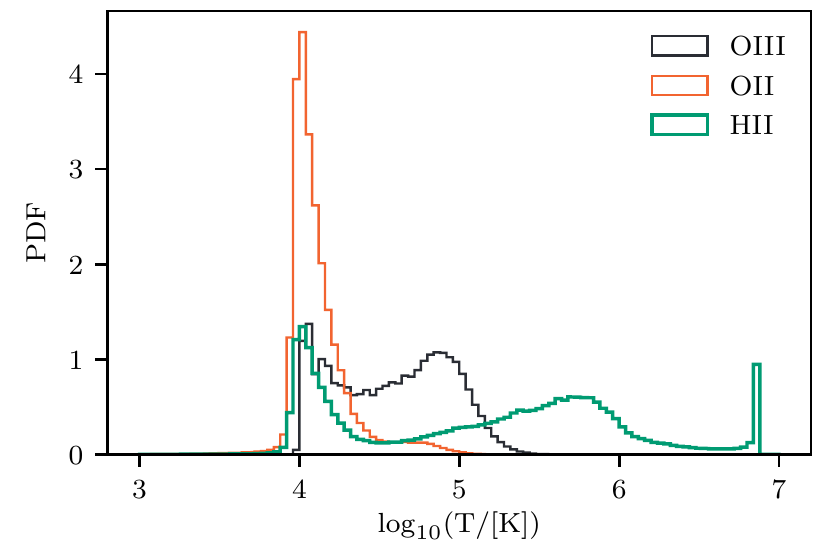}
}
\caption{(Top) Evolution of the emission line luminosities needed to apply the [O\ III] $T_e$ method as a function of time in the simulation. (Bottom) Example ion mass-weighted temperature distribution derived from the simulation at 400~Myr, showcasing the extended and non-trivial temperature distribution at a particular time.}
\label{fig:prop_evol}
\end{figure}

\begin{figure}
\centerline{
\includegraphics[scale=1,trim={0 0.0cm 0cm 0.0cm},clip]{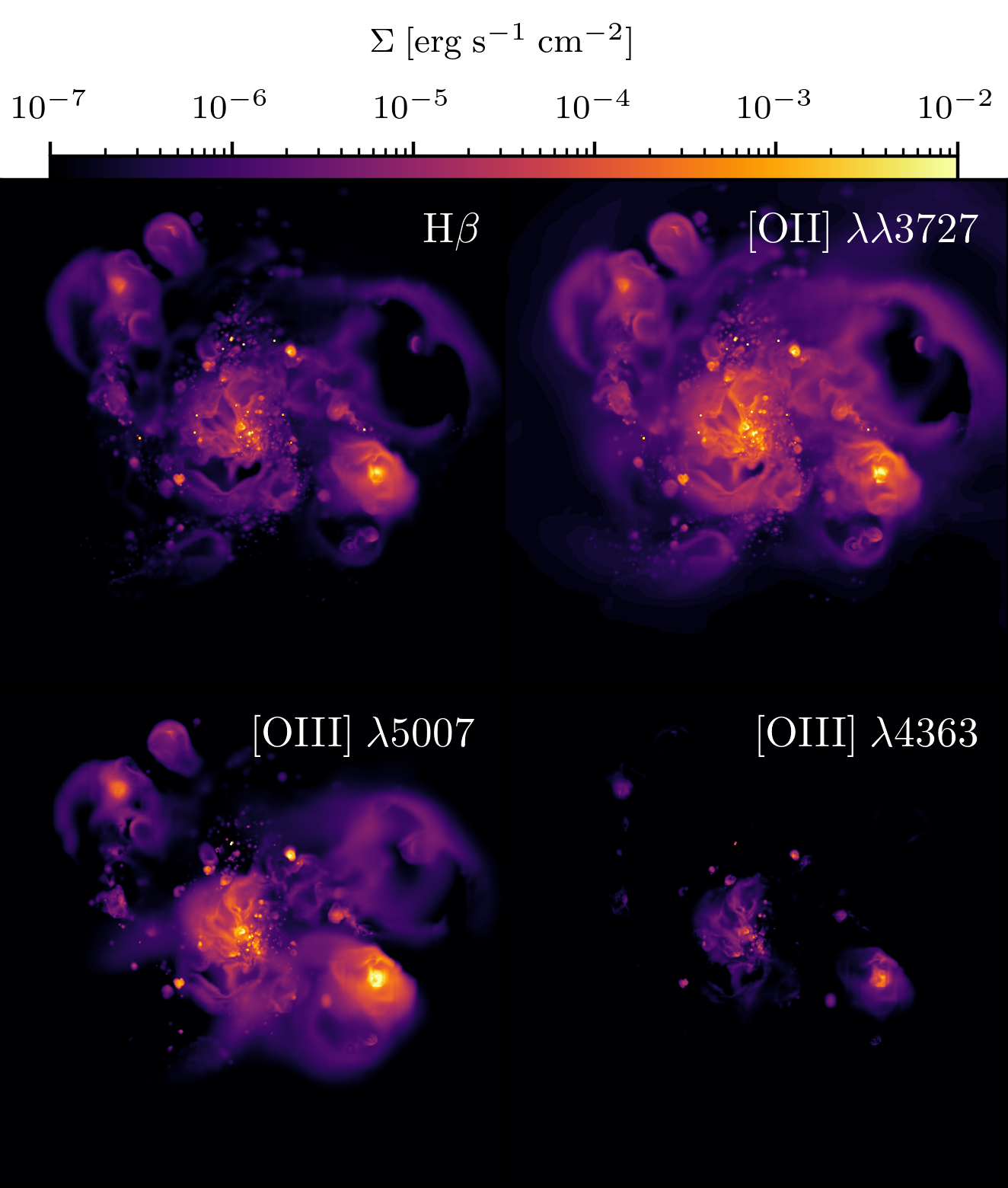}
}
\caption{4.5~kpc maps of H$\beta$, [O~{\sc ii}]~$\lambda\lambda$3727, [O~{\sc iii}]~$\lambda$5007, and [O~{\sc iii}]~$\lambda$4363, surface brightness at $t=400$~Myr, showing the spatial variations across H~{\sc ii} regions at a given time.}
\label{fig:emission_maps}
\end{figure}



\section{How temperature fluctuations affect $T_{\lowercase{e}}$-based metallicities}
\label{sec:simple}

To gain insights into the importance of sub-resolution temperature fluctuations, we construct a toy model that derives $T_e$-metallicities from O~{\sc ii} and O~{\sc iii} emission lines which are generally readily available for extragalactic studies. Ionized oxygen and hydrogen trace each other in the ISM due to their near-identical ionization potentials and a strong coupling from charge exchange reactions. Assuming in these ionized regions that the O$^{+}$ and O$^{++}$ ionization states dominate, the metallicity is:
\begin{equation}
\label{eqn:DM}
\frac{\rm O}{\rm H}\approx\frac{{\rm O}^{+}}{{\rm H}^{+}}+\frac{{\rm O}^{++}}{{\rm H}^{+}}=\frac{L_{\lambda\lambda3727}}{L_{\rm H\beta}}\frac{\epsilon_{\rm H\beta}(T_1)}{\epsilon_{\lambda\lambda3727}(T_2)}+\frac{L_{\lambda5007}}{L_{\rm H\beta}}\frac{\epsilon_{\rm H\beta}(T_3)}{\epsilon_{\lambda5007}(T_4)},
\end{equation} 
where $L_{\lambda5007}$, $L_{\lambda\lambda3727}$, and $L_{\rm H\beta}$ are the observed luminosities of the [O~{\sc iii}]~$\lambda5007$, [O~{\sc ii}]~$\lambda\lambda3727$, and H$\beta$ lines, and $\epsilon_{\lambda5007}$, $\epsilon_{\lambda\lambda3727}$, and $\epsilon_{\rm H\beta}$ are their emissivities\footnote{Formally, the emissivity is dependent on both the electron temperature ($T_e$) and density ($n_e$). This density dependence is negligible over the density range we consider in this paper ($n_e\lesssim1000$ cm$^{-3}$) and is dropped for clarity.}.

The key to this calculation is the measurement of the temperature. 
It is typical to assume $T_3=T_4$ and adopt the ratio temperature ($T_{\rm ratio}$) measured from the [O~{\sc iii}]~$\lambda4363$/[O~{\sc iii}]~$\lambda5007$ ratio.
Then $T_1=T_2$ are either measured from [O {\sc ii}] auroral lines, or derived by assuming a relation between the O {\sc ii} and O {\sc iii} temperatures \citep[e.g.][]{PM2017}.
Indeed, applying Equation~\ref{eqn:DM} with $T_{\rm ratio}$ to a single parcel of gas with uniform temperature provides an unbiased estimate of the metallicity, which depends only on atomic physics (Figure~\ref{fig:temp_sims}, left). 

However, when multiple parcels of gas with a range of temperatures are present, we can introduce the `line temperature', $T_{\rm line}$ \citep[e.g.][]{Stasinska1978}:
\begin{equation}
\label{eqn:t_line}
    \epsilon_{\lambda}(T_{\rm line}) = \frac{\int n_{X_i}n_e\epsilon_{\lambda}(T)dV}{\int n_{X_i}n_edV},
\end{equation}
where $n_e$ is the electron density, $n_{X_i}$ is the number density of species $X_i$ with a transition that gives rise to an emission line with wavelength $\lambda$, and $\epsilon_{\lambda}$ is the emissivity of that line. $T_{\rm line}$ encodes the average emissivity of each emission line across the range of probed temperatures, weighted by the local electron and ion densities.
This quantity is related to the intrinsic `ionic temperature' (the mass-weighted average temperature of a given ion), however $T_{\rm line}$ is more directly related to observed emission line fluxes as it additionally encodes information of how emissivity scales with temperature.
The crux of the bias described in Section~\ref{sec:intro} is that, since the emissivity of H$\beta$, [O~{\sc ii}]~$\lambda\lambda3727$, and [O~{\sc iii}]~$\lambda5007$ scale differently with temperature, each line has a different $T_{\rm line}$ differing from $T_{\rm ratio}$. Hence using $T_{\rm ratio}$ in Equation~\ref{eqn:DM} can result in an incorrect metallicity.

To gauge the magnitude of this effect, we construct a system comprised of 1,000 gas parcels, all with constant metallicity of $12+\log_{10}({\rm O/H})=8.48$, randomized electron densities between $10^{-1}-10^3\ {\rm cm^{-3}}$, and randomized temperatures. We construct 28 systems assuming either a normal ($\sigma=10^3$~K) or lognormal ($\sigma=0.3$~dex) temperature distribution, each with 14 different mean temperature values ranging from 7,000~K to 20,000~K. Within each parcel, the gas temperature is the same for each ion. We then obtain the total luminosities of [O~{\sc iii}]~$\lambda5007$, [O~{\sc iii}]~$\lambda4363$, [O~{\sc ii}]~$\lambda\lambda3727$, and H$\beta$ by summing the emergent luminosities from each gas parcel.\footnote{
We note that our toy model encodes no information on the spatial structure of the temperature variations as this information would be lost when the luminosity from each parcel is summed.
The cartoon in the top row of Figure~\ref{fig:temp_sims} is an arbitrary illustration of such fluctuations, but in practice the spatial distribution does not have any bearing on this experiment.
}
Applying the $T_e$-method, we compute the `measured' metallicity of the system by deriving $T_{\rm ratio}$ from [O~{\sc iii}]~$\lambda4363$/$\lambda5007$ and inputting this into Equation~\ref{eqn:DM}.
The centre panel of Figure~\ref{fig:temp_sims} shows that metallicities derived using $T_{\rm ratio}$ systematically underpredict the true value, and the magnitude of the bias strongly depends on both the shape of the temperature distribution (blue versus orange) and its mean temperature (left to right). Since mean temperature is often correlated with galaxy metallicity, temperature fluctuations can affect \textit{relative} metallicity measurements, and impact the shape of the MZR.

In contrast, if we adopt the corresponding $T_{\rm line}$ for each $T_i$ in Equation~\ref{eqn:DM}, we obtain an unbiased metallicity measurement, regardless of the underlying temperature distribution (Figure~\ref{fig:temp_sims} right panel).
However, $T_{\rm line}$ is not a directly measurable quantity. 
Given that the sub-resolution temperature distributions in real galaxies are much more complicated than the simple distributions presented here, this motivates a theoretical approach to predicting what distribution the temperatures in a realistic galaxy might take.
Thus, we now present a new, physically-motivated exploration of the temperature distribution in a simulated isolated dwarf galaxy. Based on this, we outline a novel method to convert $T_{\rm ratio}$ to $T_{\rm line}$ and debias $T_e$-derived metallicities.


\section{Quantifying temperature fluctuations in a simulated dwarf galaxy} \label{sec:results}

We now make use of novel numerical methods that allow us to directly predict the temperature distribution of H~{\sc ii} regions within a simulated galaxy, and estimate a relation between $T_{\rm ratio}$ and $T_{\rm line}$.

\vspace{-10pt} %
\subsection{Simulation methods}
\label{sec:methods}
Our simulations achieve a resolution of 4.5~pc across the ISM, resolving the structure of large H~{\sc ii} regions. Using the out-of-equilibrium temperatures, ionization states, and electron densities self-consistently produced by the simulation, we solve for equilibrium level populations to predict emission line luminosities. We follow \cite{Katz2022-mgii} to compute Balmer lines according to the most recent atomic data in {\small CHIANTI} \citep{Dere2019}, while oxygen line luminosities are calculated with {\small PyNEB} \citep{pyneb}. 
We simulate the same isolated dwarf galaxy as in \cite{RTZ} with a halo mass of $10^{10}\ {\rm M_{\odot}}$, a circular velocity of $30\ {\rm km\ s^{-1}}$, a disk gas mass of $3.5\times10^{8}\ {\rm M_{\odot}}$, and an initial central metallicity of $10^{-1}Z_{\odot}$ (see \citealt{RTZ}, Section 4 for further details). We then evolve these initial conditions using the {\small RAMSES-RTZ} code \citep{RTZ} and the {\small PRISM} interstellar medium (ISM) model for cooling and heating \citep{Prism2022}. We improve the resolution by a factor 4 compared to \cite{RTZ} to better resolve H~{\sc ii} regions, and run the simulation for twice as long.

Radiation hydrodynamics is self-consistently computed in eight energy bins (\citealt{Kimm2017}, Table 2), assuming a reduced speed of light ($c_{\rm sim}=c/100$), and is coupled to the non-equilibrium chemistry of eight ionization states of O, seven of N, six of C, and six of Si, Mg, Fe, S, and Ne, and all ionization states of H and He. We account for star formation, stellar feedback and enrichment from massive stars (\citealt{RTZ}, Section 4), and apply a \cite{Haardt2012} UV background which is exponentially suppressed at $\rho>10^{-2}\ {\rm cm^{-3}}$. We employ a fixed cosmic-ray background ionization rate of $\eta_{\rm cr}=10^{-16}\ {\rm s^{-1}H^{-1}}$. We highlight that ISM conditions are sensitive to the choice of feedback model \citep[e.g.][]{Roca2021}, and that future works are required to assess the extent of these uncertainties and how they affect the robustness of derived temperature fluctuations in simulated galaxies. Similarly, uncertainties associated with numerical metal mixing could affect the distribution of metals and their temperature distributions. In particular, adaptive mesh refinement (AMR), as used in this study, can over-mix the fluid \citep[e.g.][]{Wadsley2008} and smooth out metal (and temperature) distributions. This would, however, decrease temperature fluctuations across the galaxy, making results in this section conservative with respect to this issue.

\subsection{Abundance discrepancy in our simulated galaxy}
\label{sec:g8_discrep}

To provide the cleanest test of the $T_e$-method, we work only with the intrinsic luminosities so as to not introduce uncertainties due to dust attenuation. We assume that we detect [O~{\sc ii}]~$\lambda\lambda$3727, [O~{\sc ii}]~$\lambda\lambda$7320, 7330, [O~{\sc iii}]~$\lambda$5007, [O~{\sc iii}]~$\lambda$4363, and H$\beta$. While it is rare to simultaneously detect the [O~{\sc ii}] and [O~{\sc iii}] auroral lines in the same galaxy, this provides the cleanest test of the accuracy of the direct method.

We evolve the galaxy for 400~Myr. After the initial starburst relaxes from the initial conditions, the star formation rate (SFR) fluctuates in the range $10^{-2}<{\rm SFR/M_{\odot}yr^{-1}}<10^{-1}$. These temporal fluctuations are reflected on the galaxy-integrated emission line luminosities needed for the $T_e$-method (Figure~\ref{fig:prop_evol}, top panel). We further show in the bottom panel of Figure~\ref{fig:prop_evol} the ion mass-weighted temperature distributions for O~{\sc iii}, O~{\sc ii}, and H~{\sc ii} at the end of the simulation, and emission-line maps (Figure~\ref{fig:emission_maps}). At a given time, the dynamic ISM environment generates an extended ion temperature distribution with non-trivial shape, and an irregular emission line spatial distribution. These temperature fluctuations will cause a bias in the $T_e$-method.

\begin{figure}
\centerline{
\includegraphics[scale=1,trim={0 0.0cm 0cm 0.0cm},clip]{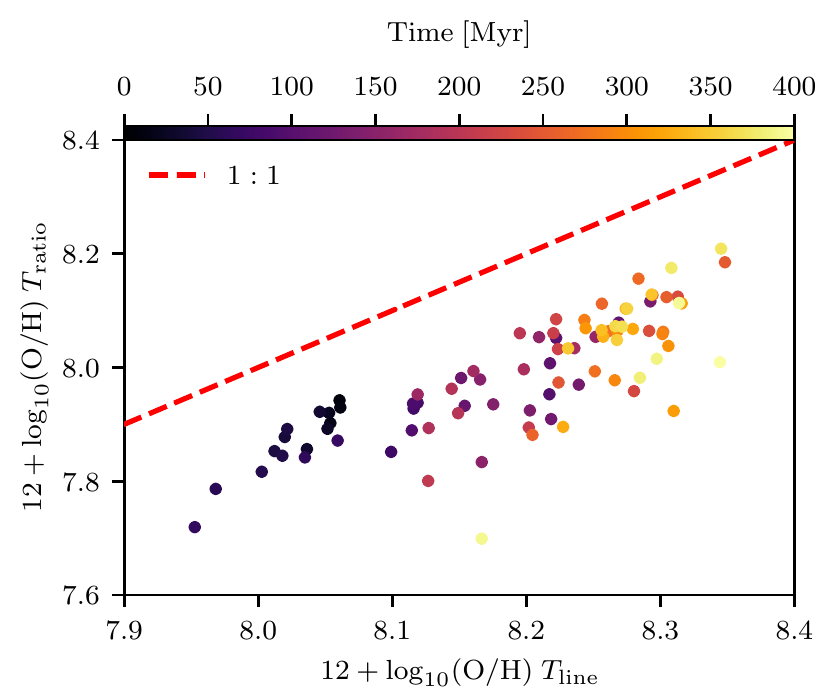}
}
\caption{Metallicity measured with $T_{\rm line}$ vs. metallicity measured with $T_{\rm ratio}$ as a function of time. As the simulation progresses (colours), the metallicity of the galaxy grows but the $T_{\rm ratio}$-derived metallicity always results in a systematically lower value.}
\label{fig:metal_predict_line_ratio}
\end{figure}

From the simulated [O~{\sc ii}]~$\lambda\lambda$3727, [O~{\sc ii}]~$\lambda\lambda$7320, 7330, [O~{\sc iii}]~$\lambda$5007, [O~{\sc iii}]~$\lambda$4363, and H$\beta$ luminosities, we then calculate metallicity using the standard direct method (Equation~\ref{eqn:DM}) using the ratio temperature derived from [O~{\sc iii}]~$\lambda4363$ / [O~{\sc iii}]~$\lambda5007$ for O~{\sc iii} and [O~{\sc ii}]~$\lambda\lambda$7320,7330 / [O~{\sc ii}]~$\lambda\lambda$3727 for O~{\sc ii}.
We further compute the unobservable $T_{\rm line}$ (Equation~\ref{eqn:t_line}) for [O~{\sc ii}]~$\lambda\lambda$3727, [O~{\sc iii}]~$\lambda$5007 and H$\beta$ using the self-consistent simulated distribution of temperatures, electron and ion densities, and compare the derived metallicities between the two in Figure~\ref{fig:metal_predict_line_ratio}. Using $T_{\rm ratio}$ results in a systematic underprediction with an average deficit of 0.21~dex, ranging between at $0.1-0.5$~dex during the course of the simulation (colours). These offsets have comparable magnitude to discrepancies between $T_e$ and RL metallicities \citep{Garcia2007}, and between galaxy-integrated measurements compared to combining their individual H~{\sc ii} regions \citep{Kobulnicky1999}.

A possible avenue to correct such abundance discrepancy is to use the $t^2$-method \citep{Peimbert1969}, in which ion temperatures are measured from multiple different tracers to better estimate their underlying temperature distribution. However, detecting multiple independent temperature tracers is highly challenging beyond the local Universe, limiting the applicability of this approach for extragalactic studies. Instead, in the next section we introduce a novel approach in which we suggest that $T_{\rm ratio}$ can be converted to $T_{\rm line}$ which, as discussed in Section~\ref{sec:simple}, results in an unbiased metallicity measurement.
We foresee that this approach will be more widely applicable to extragalactic studies where most observations have at most one temperature tracer.

\subsection{Calibrating the $T_{\rm line}$-$T_{\rm ratio}$ relation}

\begin{figure*}
\centerline{
\includegraphics[scale=1,trim={0 0.0cm 0cm 0cm},clip]{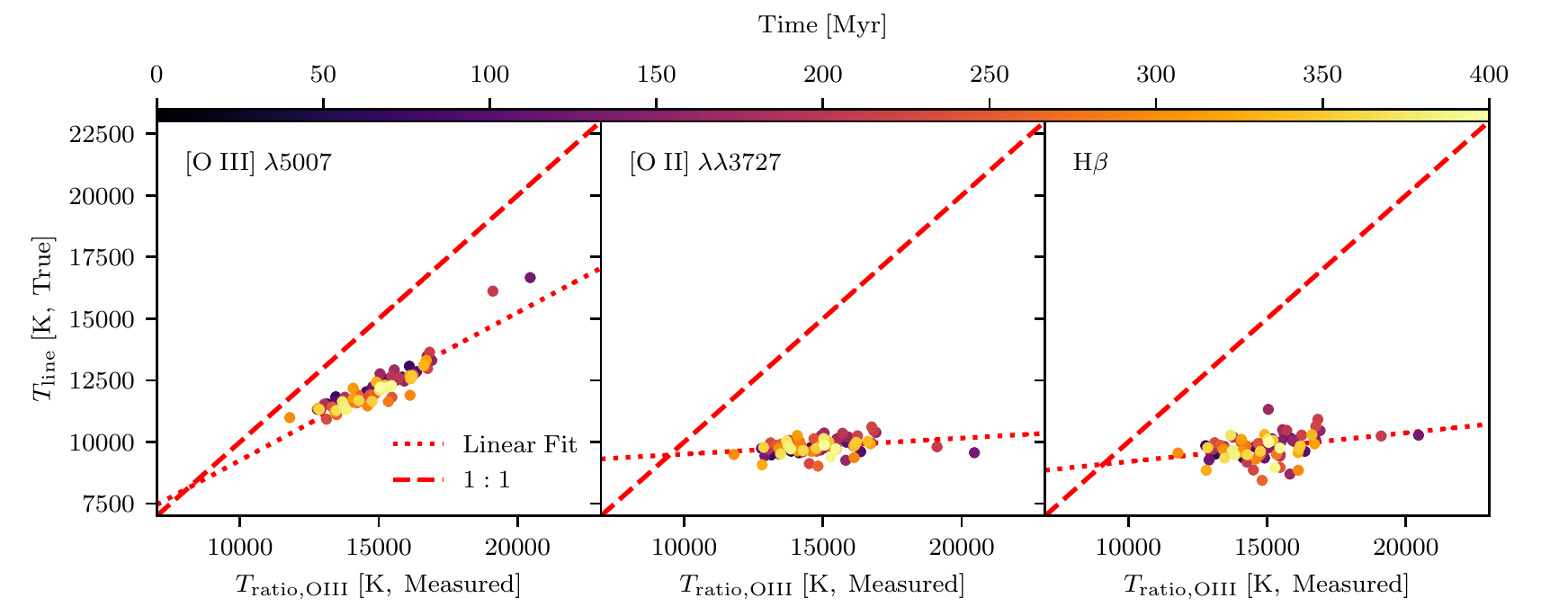}
}
\caption{$T_{\rm ratio}$ from [O {\sc iii}] $\lambda$4363/[O {\sc iii}] $\lambda$5007 compared to the [O {\sc iii}]~$\lambda$5007 (left), [O {\sc ii}]~$\lambda\lambda$3727 (centre), and H$\beta$ (right) line temperatures. All $T_{\rm line}$ are systematically underpredicted by $T_{\rm ratio}$, but they exhibit a tight correlation for each emission line.}
\label{fig:electron_t_ion_t}
\end{figure*}

Figure~\ref{fig:electron_t_ion_t} shows the ratio temperature as measured from [O~{\sc iii}]~$\lambda$4363/$\lambda$5007 versus the line temperatures of [O~{\sc iii}]~$\lambda$5007, [O~{\sc ii}]~$\lambda\lambda$3727, and H$\beta$ for each time snapshot of our simulated galaxy between 50 and 400~Myr. The temperatures of all three lines are tightly correlated with $T_{\rm ratio}$, such that 
\begin{equation}
\begin{split}    
T_{\rm line,\,[OIII]\, \lambda5007} &=0.60 \, T_{\rm ratio}+3258\ {\rm K}, \\
T_{\rm line,\,[OII]\, \lambda\lambda3727} &=0.065 \, T_{\rm ratio}+8859\ {\rm K}, \ \text{and} \\
T_{\rm line,\,H\beta} &=0.117 \, T_{\rm ratio}+8034\ {\rm K}.
\label{eqn:fits}
\end{split}
\end{equation}

We emphasise that these relations are derived from a single simulation of an isolated dwarf galaxy.
It remains unclear how these generalise to galaxies of different masses, metallicities, and star formation rates, and how sensitive they are to the choice of feedback model. 
A detailed quantification on how to generalise the coefficients in Equation~\ref{eqn:fits} across the whole galaxy population is beyond the scope of this work, and we stress that our determinations should not be applied blindly to large datasets. 
However, our simulation currently provides the most realistic estimate of temperature fluctuations across dwarf galaxies -- we thus extrapolate them to a range of stellar masses in the next section to illustrate how accounting for temperature fluctuations could reshape the low-mass end of the MZR, strongly motivating future studies extending our analysis.

We note that the relations described in Equation~\ref{eqn:fits} could also be derived in terms of $T_{\rm ratio}$(O~{\sc ii}).
Indeed, at higher masses and metallicities, the [O~{\sc ii}] auroral lines may become the most accessible temperature diagnostic \citep[e.g.][]{Sanders2022}.
However, given that the [O~{\sc iii}] $\lambda$4363 / $\lambda$5007 ratio is the most widely used temperature diagnostic in extragalactic studies (especially in the dwarf galaxy regime), we have chosen to base the relations in Equation~\ref{eqn:fits} on $T_{\rm ratio}$(O~{\sc iii}). Consideration of an O~{\sc ii} based $T_{\rm line}-T_{\rm ratio}$ relation is deferred to future work that will study higher mass and more metal-enriched galaxies.

\section{Discussion and Conclusions}
\label{sec:conclusions}
We have shown how temperature fluctuations within an H~{\sc ii} region and throughout a galaxy can bias $T_e$-method metallicities low, with a magnitude that depends on the shape and mean of the underlying distribution. While the existence of this bias is well established \citep[e.g.][]{Peimbert1967,Stasinska2005}, estimating its magnitude and correcting it in extragalactic samples is challenging (e.g. \citealt{Peimbert1967}, \citetalias{Andrews2013}). We propose a new solution based on deriving a relation between the observable $T_{\rm ratio}$ (that can be derived from auroral-to-nebular line ratios) to the unobservable $T_{\rm line}$, which we demonstrate provides a much more accurate temperature estimator for the $T_e$ method (Figure~\ref{fig:temp_sims}).

We calibrate this relation using high-resolution (4.5 pc) numerical simulations of a dwarf galaxy with {\small RAMSES-RTZ} that couple non-equilibrium metal chemistry to multi-frequency radiation hydrodynamics to estimate the temperature distributions across the warm-ionised gas in the galaxy.
Applying the standard $T_e$-method to simulated data underpredicts the representative metallicity of H~{\sc ii} regions at all times, by 0.21~dex on average and up to 0.5~dex (Figure~\ref{fig:metal_predict_line_ratio}). Nonetheless, each line temperature tightly correlates with $T_{\rm ratio}$ from [O {\sc iii}] (Figure~\ref{fig:electron_t_ion_t}), offering us the opportunity to debias metallicity estimates.

\begin{figure}
\centerline{
\includegraphics[scale=1,trim={0 0.cm 0cm 0cm},clip]{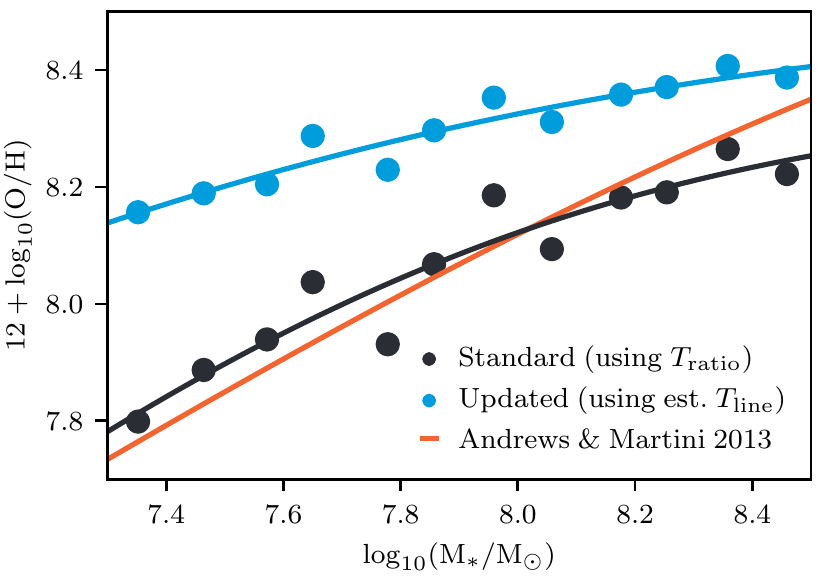}
}
\caption{MZR for SDSS galaxies based on the standard $T_e$ method using $T_{\rm ratio}$ (black), our updated $T_e$ method using the estimated $T_{\rm line}$ (blue), and the fit from \protect\citetalias{Andrews2013} (red). The slope of the MZR flattens and the normalization increases using our new approach to debiasing the measurement.}
\label{fig:mass_metallicity_recalibrated}
\end{figure}

To illustrate the importance of accounting for unresolved temperature fluctuations when determining the MZR, we use the stacked SDSS spectra from \citetalias{Andrews2013} for stellar mass bins below $10^{8.5}\ {\rm M_{\odot}}$, and apply the $T_e$-method from the [O~{\sc iii}]~$\lambda$4363, [O~{\sc iii}]~$\lambda$5007, [O~{\sc ii}]~$\lambda\lambda$3727, and H$\beta$ fluxes. Figure~\ref{fig:mass_metallicity_recalibrated} shows the resulting metallicities (black points), with the best fit (black line) using the functional form in \citetalias{Andrews2013} (their Equation~5). The original fit to the MZR from \citetalias{Andrews2013} (red line) shows minor differences due to their different atomic data, a different relation\footnote{We have adopted the relation from \cite{PM2017} assuming an electron density of 300. This provides good agreement with the original MZR from \citetalias{Andrews2013}. We emphasize that adopting different relations will fundamentally change the measured MZR.} to convert between $T_e$ (O~{\sc iii}) and $T_e$(O~{\sc ii}), and their more extended stellar mass range for the fit.

We then use Equation~\ref{eqn:fits} to estimate $T_{\rm line}$ for each emission line before deriving the metallicity from Equation~\ref{eqn:DM} and recalculate the MZR (blue points). The metallicity increases at all stellar masses, with the stellar-mass dependent offset reaching $\sim 0.3$ dex for lower mass bins, and flattens the measured slope of the MZR from 0.87 to 0.58.

Further work is needed to establish the robustness of applying Equation~\ref{eqn:fits} derived from a single dwarf galaxy to the general sample from \citetalias{Andrews2013} (as well as to other, potentially less biased, samples), however this experiment emphasizes that a careful quantification and correction of biases in metallicity measurements is essential to robustly interpret the MZR. We advocate for larger suites of comparable simulations than used in this study, across a larger range of stellar masses and metallicities, and in a cosmological context. 
Refining the relations presented in Equation~\ref{eqn:fits} from such a suite of simulations could aid interpretations of observations in a manner comparable to how 1D photoionisation models (from e.g. CLOUDY or MAPPINGS) have historically been used.

Finally, other metallicity diagnostics that are readily-available for extragalactic samples, in particular theory-calibrated strong-line methods, have shown offsets compared to metallicities derived using $T_e$-based methods (e.g. \citetalias{Andrews2013}). We plan in future work to use similar simulations as used in this work to explore the physical nature of these offsets and propose new corrections to improve the robustness of metallicity estimates.

\section*{Acknowledgements}
This work was performed using the DiRAC Data Intensive service at Leicester, operated by the University of Leicester IT Services, which forms part of the STFC DiRAC HPC Facility. The equipment was funded by BEIS capital funding via STFC capital grants ST/K000373/1 and ST/R002363/1 and STFC DiRAC Operations grant ST/R001014/1. DiRAC is part of the National e-Infrastructure.
AJC acknowledges funding from the `FirstGalaxies' Advanced Grant from the European Research Council (ERC) under the European Union's Horizon 2020 research and innovation programme (Grant agreement No. 789056).
MR is supported by the Beecroft Fellowship funded by Adrian Beecroft.

\section*{Data Availability}
Underlying data will be shared on reasonable request to the authors.


\bibliographystyle{mnras}
\bibliography{example} 

\bsp	
\label{lastpage}
\end{document}